\documentclass[11pt]{article}
\usepackage{amsmath}
\usepackage{amssymb}
\usepackage{amsfonts}
\usepackage[all,poly]{xy}

\newcounter{algoctr}

\newif\ifnotesw\noteswtrue
   {\ifnotesw\marginpar[\hfill\(\top\)]{\(\top\)}\fi}%
   {\ifnotesw\marginpar[\hfill\(\bot\)]{\(\bot\)}\fi}

\newcommand{\mnote}[1]%
    {\ifnotesw\marginpar%
        [{\scriptsize\begin{minipage}[t]{\marginparwidth}
        \raggedleft#1%
                        \end{minipage}}]%
        {\scriptsize\begin{minipage}[t]{\marginparwidth}
        \raggedright#1%
                        \end{minipage}}%
    \fi}

\newcommand{\ignore}[1]{}

\newsavebox{\given}
\savebox{\given}[1em]{\rule[-1.5ex]{.2mm}{4ex}}

\newcommand{\bnum}{\begin{equation}}
\newcommand{\enum}{\end{equation}}

\newtheorem{theorem}{Theorem}
\newtheorem{corollary}[theorem]{Corollary}
\newtheorem{lemma}[theorem]{Lemma}

\newtheorem{conjecture}[theorem]{Conjecture}

\newtheorem{definition}{Definition}

\newcommand{\blackslug}{\rule{7pt}{7pt}}

\newcommand{\iverson}[1]{\lbrack\!\lbrack #1 \rbrack\!\rbrack}

\newcommand{\real}{\ifmmode {\rm R} \else ${\rm R}$ \fi}
\newcommand{\nat}{\ifmmode {\rm N} \else ${\rm N}$  \fi}
\newcommand{\tot}{\ifmmode {\cal T} \else ${\cal T}$ \fi}
\newcommand{\sigstar}{\ifmmode \Sigma^{\ast} \else $\Sigma^{\ast}$ \fi}

\newcommand{\inn}{\ifmmode \in \else $\in$ \fi}
\renewcommand{\phi}{\ifmmode \varphi \else $\varphi$ \fi}
\renewcommand{\le}{\ifmmode \leq \else $\leq$ \fi}
\renewcommand{\ge}{\ifmmode \geq \else $\geq$ \fi}
\renewcommand{\ne}{\ifmmode \neq \else $\neq$ \fi}
\newcommand{\lt}{\ifmmode < \else $<$ \fi}
\newcommand{\gt}{\ifmmode > \else $>$ \fi}
\newcommand{\eq}{\ifmmode = \else $=$ \fi}
\newcommand{\half}{\ifmmode \frac{1}{2} \else $\frac{1}{2}$ \fi}
\newcommand{\oneovern}{\ifmmode \frac{1}{n} \else $\frac{1}{n}$ \fi}

\newcommand{\ra}{\ifmmode \rightarrow \else $\rightarrow$ \fi}
\newcommand{\qed}{\hfill{\setlength{\fboxsep}{0pt}
\framebox[7pt]{\rule{0pt}{7pt}}}}

\renewcommand{\notin}{\ifmmode \not\in \else $\not\in$ \fi}
\newlength{\thislabel}
\newcommand{\labsize}[1]{\settowidth{\thislabel}{#1}}

\newcommand{\prf}{\par\noindent{\sl Proof } \hspace{.01 in}}

\newcommand{\lip}{\langle}
\newcommand{\rip}{\rangle}
\def\Nat{\mathbb N}
\def\Complex{\mathbb C}
\def\Int{\mathbb Z}

\newcommand{\ket}[1]{| #1 \rip}
\newcommand{\braket}[2]{\lip #1 | #2 \rip}

\newcommand{\dt}{\mathsf{d}t}

\newcommand{\crc}[1]{\lip #1 \rip}

\title{
Mixing of Quantum Walk on Circulant Bunkbeds\footnote{Supported in part by NSF grants DMR-0121146 and DMS-0353050.}
} 
\author{
{Peter Lo}\\{\em St. Mary's College}
\and {Siddharth Rajaram}\\{\em Middlebury College} 
\and {Diana Schepens}\\{\em Houghton College}
\and {Daniel Sullivan}\\{\em Swarthmore College}
\and {Christino Tamon}\footnote{Contact author: tino@clarkson.edu}\\{\em Clarkson University}
\and {Jeffrey Ward}\\{\em Clarkson University}
}
\date{\today}

\begin{document}
\bibliographystyle{plain}
\maketitle

\begin{abstract}
\ignore{
We study a continuous-time quantum walk on circulants and their bunkbed extensions obtained via 
the standard join $G + H$ and Cartesian product $G \oplus H$ binary operators of graphs $G$ and $H$.
First, we show that the quantum walk is uniform mixing on circulants with bounded eigenvalue multiplicity. 
This extends the known fact about the uniform mixing of cycles.
Then, we analyze a quantum walk on the join of circulants. Our analysis on the join of two circulants sheds
some light on why a quantum walk on the complete graph is not uniform mixing. This is obtained by viewing 
$K_{n}$ as the join $K_{1} + K_{n-1}$. Our analysis on the homogeneous join $G + \ldots + G$ of a circulant
$G$ reveals the non-uniform mixing of the complete multipartite graphs.
Finally, we analyze the quantum walk on the Cartesian product of a path and a circulant.
Our analysis on $P_{2} \oplus C$ highlights a difference between $\Int_{n}$-circulants and $\Int_{2}^{n}$-circulants
(which include the hypercubes).
Our proofs employ purely elementary arguments based on the spectra of the underlying graphs.
}
We give new observations on the mixing dynamics of a continuous-time quantum walk on circulants 
and their bunkbed extensions. These bunkbeds are defined through two standard graph operators:
the {\em join} $G + H$ and the {\em Cartesian product} $G \oplus H$ of graphs $G$ and $H$. 
Our results include the following:
\begin{itemize}
\item The quantum walk is average uniform mixing on circulants with bounded eigenvalue multiplicity. 
	This extends a known fact about the cycles $C_{n}$.
\item Explicit analysis of the probability distribution of the quantum walk on the join of circulants.
	This explains why complete partite graphs are not average uniform mixing, 
	using the fact $K_{n} = K_{1} + K_{n-1}$ and $K_{n,\ldots,n} = \overline{K}_{n} + \ldots + \overline{K}_{n}$.
\item The quantum walk on the Cartesian product of a $m$-vertex path $P_{m}$ and a circulant $G$,
	namely, $P_{m} \oplus G$, is average uniform mixing if $G$ is.
	This highlights a difference between circulants and the hypercubes $Q_{n} = P_{2} \oplus Q_{n-1}$.
\end{itemize}
Our proofs employ purely elementary arguments based on the spectra of the graphs. 

\vspace{.1in}

\end{abstract}

\section{Introduction}

The study of continuous-time quantum walks on graphs has important potential applications in quantum computation
\cite{nc00}.  First, as an algorithmic technique, it was used to devise efficient quantum search algorithms with 
considerable speedup over classical algorithms \cite{ccdfgs03}. Second, it may provide a simpler physical 
implementation of a quantum computer, given that there is an abundance of physical processes that simulate
quantum walk on graphs \cite{drkb02}. In the physics literature, continuous-time quantum walks is mainly studied 
over infinite constant-dimensional lattices, such as the one-dimensional line (see \cite{fls}, Chapters 13,16). 
On the other hand, the study of random walks on general graphs is a topic of broad interest in the mathematics 
and computer science community \cite{bollobas,l93}.

In this paper, we study the mixing dynamics of continuous-time quantum walks on circulant graphs. More particularly, 
we consider the {\em average} or limiting probability distribution of a quantum walk. This notion was introduced in 
\cite{aakv01} and is the quantum analogue of a stationary distribution of classical random walks. On the circulant 
graphs, our goal was to characterize the graphs for which the continuous-time quantum walk reaches (almost) uniform 
average probability distribution. It was previously known that cycles are near uniform mixing, whereas the complete 
graphs and hypercubes are not (see \cite{aaht03,mr02}). Our other goal in this paper is to discover graph theoretic
structures that may explain this polarized phenomena.

First, we show that circulants with bounded eigenvalue multiplicity are almost uniform mixing. This generalization
explains why cycles are uniform mixing. Second, we consider bunkbed graphs constructed using the join and the 
Cartesian product operators. By analyzing the join of two circulants, we observe an interesting mixing phenomena 
on the cone $K_{1} + G$ of a circulant $G$, that is dependent on the density of $G$. If the quantum walk starts
on $K_{1}$, a dense graph $G$ {\em repels} the probability away from the copy of $G$. This explains why the limiting
distribution of a quantum walk on the complete graph is not near the uniform distribution. We extend this investigation
to the homogeneous join of circulants, namely, $G + \ldots + G$, for a circulant $G$. We show that this bunkbed graph
is uniform mixing if $G$ is uniform mixing and the join is over a constant number of copies of $G$. A corollary of 
this transference property explains the non-uniform mixing of the complete multipartite graphs 
$\overline{K}_{n} + \ldots + \overline{K}_{n}$.

We also analyze bunkbed graphs obtained from the Cartesian product $P_{m} \oplus G$ of a path $P_{m}$ and a circulant 
$G$. On this bunkbed structure, we observe another transference property: the quantum walk on $P_{m} \oplus G$ is
uniform mixing if it is uniform mixing on $G$ and the path is of constant size. This highlights a striking difference
with the hypercube $Q_{n}$, since the hypercube is also a bunkbed $Q_{n} = P_{2} \oplus Q_{n-1}$, but it is known that
they are not uniform mixing \cite{mr02}. It is interesting to note that both classes of graphs are group-theoretic 
circulants (see \cite{d90}), since our circulants are the $\Int_{n}$-circulants while hypercubes are the
$(\Int_{2})^{n}$-circulants. 
This suggests a group theoretic investigation into the mixing phenomena of generalized circulants, which we leave for 
future work.

In this paper, we focus exclusively on continuous-time quantum walks. We refer the reader to \cite{kendon} for a
survey of other models of quantum walks. As a final remark, we mention that most of the graphs we consider have the 
standard stationary distributions in the classical random walks where the limiting probability of a vertex is 
proportional to its degree \cite{aldous-fill}.
\ignore{
Also, if appropriate, we will compare the non-classical mixing dynamics of 
our quantum walk with the well-known random walks on graphs \cite{aldous-fill}.
}

\section{Preliminaries}

\ignore{
{\em Notation} For a Boolean or logical statement $\xi$, let $\iverson{\xi}$ be $1$ if the statement
$\xi$ is true, and $0$ otherwise. For a natural $n \in \Nat$, let $\Int_{n}$ denote the integers $0,\ldots,n-1$
under addition modulo $n$. Let $i = \sqrt{-1}$ throughout. 
\ignore{Given two probability distributions $P$ and $Q$,
the {\em total variation} distance between them is defined as $||P - Q|| = \sum_{x} |P(x) - Q(x)|$. }
}
We consider simple, undirected graphs that are connected, and mostly regular. 
For a graph $G=(V,E)$, let $A_{G}$ be the adjacency matrix of $G$, where $A_{G}[j,k] = \iverson{(j,k) \in E}$. 
Here and throughout, we will use $\iverson{\Psi}$ to denote the characteristic function of a logical statement
$\Psi$, that is, $1$ if $\Psi$ is true, and $0$ if it is false.
The set of eigenvalues of $A_{G}$ is denoted $Sp(G)$, and the (algebraic) multiplicity of an eigenvalue 
$\lambda$ is denoted $m(\lambda)$.
The spectral {\em type} $\tau(G)$ of a graph $G$ is the number of distinct eigenvalues of the adjacency
matrix $A_{G}$ of $G$. We will denote the maximum (algebraic) multiplicity of any eigenvalue of graph $G$ 
by $\mu(G)$.
Some of the families of graphs that we will consider include 
the complete multipartite graphs $K^{(m)}_{n}$, where there are $m$ partitions with a partition size of $n$,
the cycles $C_{n}$ and paths $P_{n}$, and the hypercubes $Q_{n}$. 
Relevant background on graphs and their spectral properties can be found in \cite{biggs}.

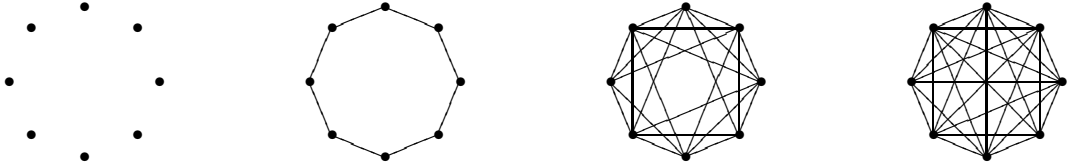
\begin{figure}[t]
\[\begin{xy} /r10mm/:
 ,{\xypolygon8"E"{~<{}~>{}~={0}\dir{*}}}
 ,+/r40mm/
 ,{\xypolygon8"C"{~={0}\dir{*}}}
 ,+/r40mm/
 ,{\xypolygon8"S"{~={0}\dir{*}}}
 ,"S1";"S3"**@{-},"S2";"S4"**@{-},"S3";"S5"**@{-},"S4";"S6"**@{-},"S5";"S7"**@{-},"S6";"S8"**@{-},"S7";"S1"**@{-},"S8";"S2"**@{-}
 ,"S1";"S4"**@{-},"S2";"S5"**@{-},"S3";"S6"**@{-},"S4";"S7"**@{-},"S6";"S1"**@{-},"S7";"S2"**@{-},"S8";"S3"**@{-}
 ,+/r40mm/
 ,-(0.0,1.0)
 ,{\xypolygon8"K"{~={0}\dir{*}}}
 ,"K1";"K3"**@{-},"K2";"K4"**@{-},"K3";"K5"**@{-},"K4";"K6"**@{-},"K5";"K7"**@{-},"K6";"K8"**@{-},"K7";"K1"**@{-},"K8";"K2"**@{-}
 ,"K1";"K4"**@{-},"K2";"K5"**@{-},"K3";"K6"**@{-},"K4";"K7"**@{-},"K6";"K1"**@{-},"K7";"K2"**@{-},"K8";"K3"**@{-}
 ,"K1";"K5"**@{-},"K2";"K6"**@{-},"K3";"K7"**@{-},"K4";"K8"**@{-}
\end{xy}\]
\caption{Examples of circulants of order $8$.
From left to right: 
(i) the empty graph $\overline{K}_{8}$.
(ii) the cycle $C_{8}$.
(iii) a strongly-regular circulant: clique minus a perfect matching.
(iv) the complete graph $K_{8}$.
}
\label{figure:circulants}
\end{figure}

A graph $G$ is called {\em circulant} if its adjacency matrix $A_{G}$ is circulant. 
A circulant matrix $A$ is specified by its first row, say $(a_{0}, a_{1}, \ldots, a_{n-1})$, and 
is defined as $A_{j,k} = a_{k-j \pmod{n}}$, where $j,k \in \Int_{n}$. Here $\Int_{n}$ denotes the group
of integers $\{0,\ldots,n-1\}$ under addition modulo $n$.
Note that $a_{0} = 0$, since our graphs are simple, and $a_{j} = a_{n-j}$, since our graphs are undirected.
Connectivity is guaranteed if the greatest common divisor of $n$ and all indices $k$, for which $a_{k}=1$, is one.
Alternatively, a circulant graph $G = (V,E)$ can be specified by a subset $S \subseteq \Int_{n}$, where $(j,k) \in E$
if $k-j \in S$. In this case, we write $G = \crc{S}$. We will assume that $S$ is closed under taking inverses,
namely, if $d \in S$, then $-d \in S$.
Figure \ref{figure:circulants} contains some examples of circulant graphs.

It is known that circulant graphs $G$ are diagonalizable by the Fourier matrix $F$ defined as
$F_{j,k} = n^{-1/2} \omega_{n}^{jk}$, where $\omega_{n} = \exp(2\pi i/n)$. In fact, the eigenvalues of $A_{G}$ are
\begin{equation} \label{eqn:eigenvalue}
\lambda_{j} = \sum_{k=1}^{n-1} a_{k}\omega^{jk} 
	= \sum_{k=1}^{\lfloor (n-1)/2 \rfloor} 2\cos\left(\frac{2\pi jk}{n}\right) \
		+ \ \iverson{n \mbox{ even}} \ a_{n/2} \ (-1)^{j}.
\end{equation}

A {\em continuous-time quantum walk} on a graph $G=(V,E)$ is defined using the Schr\"{o}dinger equation 
with the real symmetric matrix $A_{G}$ as the Hamiltonian (see \cite{ccdfgs03}). 
If $\ket{\psi(t)} \in \Complex^{|V|}$ is a time-dependent amplitude vector on the vertices of $G$, 
then the evolution of the quantum walk is given by
\begin{equation}
\ket{\psi(t)} = e^{-it A_{G}} \ket{\psi(0)},	
\end{equation}
where $i = \sqrt{-1}$ and $\ket{\psi(0)}$ is the initial amplitude vector. We usually assume that $\ket{\psi(0)}$ 
is a unit vector, with $\braket{x}{\psi(0)} = \iverson{x = 0}$, for some vertex $0$.
The amplitude of the quantum walk on vertex $j$ at time $t$ is given by $\psi_{j}(t) = \braket{j}{\psi(t)}$, 
while the probability of being on vertex $j$ at time $t$ is $p_{j}(t) = |\psi_{j}(t)|^{2}$. 
The average (or limiting) probability of being on vertex $j$ is defined as 
\begin{equation}
\overline{p}_{j} = \lim_{T \rightarrow \infty} \frac{1}{T} \int_{0}^{T} p_{j}(t) \ \dt.
\end{equation}
This notion appeared in \cite{aakv01} in the context of discrete-time quantum walks.
The limiting probability distribution of the quantum walk will be denoted $\overline{P}$.

\begin{definition} (Average Uniform Mixing) \\
The average mixing of a continuous-time quantum walk on a graph $G=(V,E)$ is called {\em uniform} if 
$\overline{p}_{j} = O(1/|V|)$,
for each vertex $j$ of $G$.
\end{definition}

\par\noindent{\em Remark} Note that in the above definition, we only require that each limiting probability 
be linearly proportional to the uniform probability value. This is less stringent than requiring that the 
quantum walk achieves {\em exactly} uniform probability distribution (see \cite{abtw03,aaht03}).
When the graph $G$ is not regular, the limiting probability distribution $\overline{P}$ 
may depend on the initial state $\ket{\psi(0)}$. We will specify carefully the effect of the initial 
states in these cases, but suppress this dependence for vertex-transitive graphs.


\section{Mixing and Bounded Multiplicities}

\begin{theorem} \label{thm:bounded-mix}
Let $G$ be a circulant graph. If $\mu(G)$ is bounded, then the continuous-time quantum walk on $G$ is
average uniform mixing.
\end{theorem}
\prf Let $n$ be the order of $G$ and let $A$ be the adjacency matrix of $G$. 
Since $\ket{0} = \sum_{k=0}^{n-1} \frac{1}{\sqrt{n}} \ket{F_{k}}$, if $\ket{\psi(0)} = \ket{0}$, we have
$\ket{\psi(t)} = e^{-itA} \ket{\psi(0)} = \frac{1}{\sqrt{n}} \sum_{k=0}^{n-1} e^{-it \lambda_{k}} \ket{F_{k}}$.
This yields $\braket{j}{\psi(t)} = \frac{1}{n} \sum_{k=0}^{n-1} e^{-it \lambda_{k}} \omega^{jk}$. Thus,
\begin{equation}
p_{j}(t) 
	= \frac{1}{n^2} \sum_{k,\ell} e^{-it(\lambda_{k} - \lambda_{\ell})} \omega^{j(k-\ell)} 
	= \frac{1}{n} + \frac{1}{n^2} \sum_{k \neq \ell} e^{-it(\lambda_{k} - \lambda_{\ell})} \omega^{j(k-\ell)}.
\end{equation}
Using the above, the average (limiting) probabilities are
\begin{equation}
\left|p_{j} - \frac{1}{n}\right| 
	\ \le \ \frac{1}{n^2} \sum_{\lambda \in Sp(A)} \binom{m(\lambda)}{2}
	\ \le \ \frac{1}{n} \binom{\mu(G)}{2}.
\end{equation}
So, if $\mu(G) = O(1)$, we have uniform mixing.
\qed

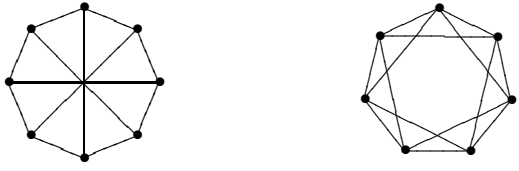
\begin{figure}[t]
\[\begin{xy} /r10mm/:
 ,{\xypolygon8"W"{~={0}\dir{*}}}
 ,"W1";"W5"**@{-},"W2";"W6"**@{-},"W3";"W7"**@{-},"W4";"W8"**@{-}
 ,+/r40mm/
 ,+(0.0,0.7)
 ,{\xypolygon7"D"{~={38}\dir{*}}}
 ,"D1";"D3"**@{-},"D2";"D4"**@{-},"D3";"D5"**@{-},"D4";"D6"**@{-}
 ,"D5";"D7"**@{-},"D6";"D1"**@{-},"D7";"D2"**@{-}
\end{xy}\]
\caption{Some circulants with bounded eigenvalue multiplicity.
From left to right: 
(i) the wheel $V_{8}$: a cycle plus a perfect matching.
(ii) a double-loop circulant $\langle \pm 1,\pm 2\rangle$ of order $7$.
}
\label{figure:maximal-circulant}
\end{figure}

\begin{theorem}
The continuous-time quantum walk on 
a constant-degree $n$-vertex circulant of the form $\crc{1, n/k_{1}, \ldots, n/k_{d}}$,
where, for each $j=1,\ldots,d$, $k_{j}$ is a constant which divides $n$,
is average uniform mixing.
\end{theorem}
\prf
\ignore{
{\em (i)} Consider $G = \crc{1,n/2}$, where $n$ is even. From Equation (\ref{eqn:eigenvalue}), we have
$\lambda_{0} = 3$ and
\begin{equation}
\lambda_{j} = 2\cos\left(\frac{2\pi j}{n}\right) + (-1)^{j}, \ \ \ j = 1,\ldots,n-1.
\end{equation}
For $j,k \in \{1,\ldots,n-1\}$, if $j \not\equiv \pm k \pmod{n}$, then $\lambda_{j} = \lambda_{k}$ only if
$j$ and $k$ differ in parities (one is odd and the other is even). This shows that no three such eigenvalues
can share the same value, since two of them must have the same parity. Thus, $\mu(G) \le 3$.
}
The eigenvalues of $G$ are given by
\begin{equation}
\lambda_{j} 
	= 2\cos\left(\frac{2\pi j}{n}\right) 
	+ 2 \sum_{\ell = 1}^{d} \cos\left(\frac{2\pi j}{k_{\ell}}\right),
\end{equation}
for $j=1,\ldots,n-1$. 
Since the sum $\sum_{\ell=1}^{d} \cos(2\pi j/k_{\ell})$ can have at most $\prod_{\ell=1}^{d} k_{\ell} = O(1)$ 
distinct values, each eigenvalue must have a constant multiplicity. By Theorem \ref{thm:bounded-mix}, we have
the claimed result.
\ignore{
{\em (ii)} 
We outline the proof of a simpler result. 
Consider the circulant $G = \crc{1, n/4}$. Then,
\begin{equation}
\lambda_{j} 
	= 2\cos\left(\frac{2\pi j}{n}\right) + 2\cos\left(\frac{j \pi}{2}\right),
\end{equation}
where $j = 1,\ldots,n-1$. Since $\cos(j \pi/2) \in \{0,\pm 1\}$, each eigenvalue (or horizontal line) can 
appear (intersect) the shifted $2\cos(x)$ curve in at most a constant amount of places (actually, $6$).
}
\qed

\begin{corollary}
The continuous-time quantum walk on the $3$-regular circulant "wheel" $V_{n} = \crc{1,n/2}$ of even order $n$ 
is uniform mixing.
\end{corollary}


\section{Mixing on Join Bunkbeds}

In this section, we study a circulant bunkbed structure obtained by the join of circulants. 
Formally, the {\em join} $G + H$ of two graphs $G$ and $H$ is defined as to satisfy
$\overline{G + H} = \overline{G} \cup \overline{H}$ (see \cite{sw78}). It is easy to see that this is 
a graph obtained by connecting each vertex of $G$ to each vertex of $H$, while maintaining the internal 
structures of $G$ and $H$. For a graph $G$, the {\em cone} of $G$ will denote the graph $K_{1} + G$.

\begin{lemma} \label{lemma:sw}
Let $G$ and $H$ be circulants of degrees $k$ and $\ell$, respectively. 
Suppose that the eigenvalues of $G$ and $H$ are $k = \mu_{0} > \mu_{1} \ge \ldots \ge \mu_{|G|-1}$ 
and $\ell = \nu_{0} > \nu_{1} \ge \ldots \ge \nu_{|H|-1}$, respectively.
Then, the eigenvalues and (orthonormal) eigenvectors of $G + H$ are found in three separate sets 
$\{\langle \mu_{a}, \ket{z^{G}_{a}}\rangle \ : \ 1 \le a \le |G|-1\}$, 
$\{\langle \nu_{b}, \ket{z^{H}_{b}}\rangle \ : \ 1 \le b \le |H|-1\}$, 
and $\{\langle \lambda_{\pm}, \ket{z_{\pm}}\rangle\}$, where, for $x = 0,\ldots,|G||H|-1$, we have
\begin{eqnarray}
\braket{x}{z^{G}_{a}} & = & \frac{1}{\sqrt{|G|}} \ \omega_{|G|}^{ax} \ \iverson{x \in G},	\ \ a = 1,\ldots,|G|-1 \\ 
\braket{x}{z^{H}_{b}} & = & \frac{1}{\sqrt{|H|}} \ \omega_{|H|}^{bx} \ \iverson{x \in H},	\ \ b = 1,\ldots,|H|-1 \\ 
\braket{x}{z_{\pm}}   & = & \frac{1}{L_{\pm}} \ (\beta_{\pm})^{\iverson{x \in H}},
\end{eqnarray}
where $\beta_{\pm} = (\lambda_{\pm} - k)/|H|$, $L_{\pm} = \sqrt{|G| + |H|\beta_{\pm}^{2}}$, and
$\lambda_{\pm}$ are the roots of $\lambda^2 - (k + \ell)\lambda - (|G||H| - k\ell) = 0$. 
\end{lemma}
\ignore{
\begin{eqnarray}
\mu_{a} 	& \Longleftrightarrow & [\vec{g}_{a} \ \ | \ \ \vec{0}_{n}]^{T},	\ \ \ a = 1,\ldots,|G|-1 \\ 
\nu_{b} 	& \Longleftrightarrow & [\vec{0}_{m} \ \ | \ \ \vec{h}_{b}]^{T},	\ \ \ b = 1,\ldots,|H|-1 \\ 
\lambda_{\pm}  & \Longleftrightarrow & [\vec{1}_{m} \ \ | \ \ \vec{(\beta_{\pm})}_{n}]^{T}
\end{eqnarray}}
\prf
Note that the adjacency matrix of $G + H$ is given by
\begin{equation}
A = \begin{bmatrix} A_{G} & J_{|G| \times |H|} \\ J_{|H| \times |G|} & A_{H} \end{bmatrix}
\end{equation}
It is easy to see that $\ket{z^{G}_{a}}$ are eigenvectors of $A$ with eigenvalues $\mu_{a}$, for $a = 1,\ldots,|G|-1$,
and $\ket{z^{H}_{b}}$ are eigenvectors of $A$ with eigenvalues $\nu_{b}$, for $b = 1,\ldots,|H|-1$. The last two
eigenvectors are obtained by noting that the eigenvectors have the form
$\begin{bmatrix} a & \ldots & a & b & \ldots & b \end{bmatrix}^{T}$.
This gives the equations $ka + b|H| = \lambda a$ and $\ell b + a|G| = \lambda b$,
whose solutions yield the eigenvalues $\lambda_{\pm} = \frac{1}{2}((k+\ell)^{2} \pm \sqrt{\Delta})$,
where $\Delta = (k - \ell)^{2} + 4|G||H|$, and eigenvectors with $a = 1$ and $b = (\lambda_{\pm} - k)/|H|$.
\qed

\begin{theorem} \label{thm:join}
Suppose that $G$ and $H$ are circulants of degrees $k$ and $\ell$, respectively. 
Let $\Delta = (k-\ell)^{2} + 4|G||H|$ and $\lambda_{\pm} = 1/2[(k+\ell)^{2} \pm \sqrt{\Delta}]$.
Consider a continuous-time quantum walk on $G + H$ starting at some vertex of $G$. 
Let $\overline{p}_{x}(G)$ denote the limiting probability of $x \in G$ over the subgraph $G$.
Assume that 
\begin{equation}
\lambda_{-} \not\in (Sp(G) \setminus \{k\}) \cup (Sp(H) \setminus \{\ell\}).
\end{equation}
Then, the limiting probabilities of the vertices of $G + H$ are 
\begin{equation}
\overline{p}_{x}(G + H) \ = \
\left\{\left(\overline{p}_{x}(G) - \frac{1}{|G|}\right) + \frac{1}{|G|}\left(\frac{1}{|G|} - \frac{2|H|}{\Delta}\right)\right\} 
\iverson{x \in G} \ + \ \frac{2}{\Delta} \iverson{x \in H}
\end{equation}
\end{theorem}
\ignore{
	\left\{
	\begin{array}{ll}
	\left(\overline{p}_{x}(G) - \frac{1}{|G|}\right) + \frac{1}{|G|}\left(\frac{1}{|G|} - \frac{2|H|}{\Delta}\right)
		& \mbox{ if $x \in G$ }	\\ 
	\frac{2}{\Delta} & \mbox{ if $x \in H$ }
	\end{array}\right.
}
\prf
Let the initial state be $\ket{\psi(0)} = \ket{0}$ where the quantum walk starts at a vertex of $G$.
By Lemma \ref{lemma:sw}, we have
\begin{equation}
\ket{\psi(0)} 
	= \frac{1}{\sqrt{|G|}} \sum_{a=1}^{|G|-1} \ket{z^{G}_{a}} + \sum_{\pm} \frac{1}{L_{\pm}} \ket{z_{\pm}},
\end{equation}
and, thus,
\begin{equation}
\ket{\psi(t)} 
	= \frac{1}{\sqrt{|G|}} \sum_{a=1}^{|G|-1} e^{-it\mu_{a}} \ket{z^{G}_{a}} + 
		\sum_{\pm} \frac{e^{-it\lambda_{\pm}}}{L_{\pm}} \ket{z_{\pm}}.
\end{equation}
The amplitude on vertex $x$ at time $t$ is given by
\begin{equation}
\braket{x}{\psi(t)} 
	= \frac{1}{|G|}\sum_{a=1}^{|G|-1} e^{-it \mu_{a}}\omega_{|G|}^{ax} + 
		\sum_{\pm} \frac{e^{-it \lambda_{\pm}}}{L_{\pm}^{2}} \beta_{\pm}^{\iverson{x \in H}},
\end{equation}
where $\beta_{\pm} = (\lambda_{\pm} - k)/|H|$, and we obtain
\begin{equation}
\overline{p}_{x} 
	= \lim_{T \rightarrow \infty} \frac{1}{T} \int_{0}^{T} \dt \ |\braket{x}{\psi(t)}|^{2}
	= \overline{p}_{x}(G) - \frac{1}{|G|} 
		+ \sum_{\pm} \left(\frac{\beta_{\pm}^{\iverson{x \in H}}}{L_{\pm}^{2}}\right)^{2},
\end{equation}
where $\overline{p}_{x}(G)$ is the limiting probability on the subgraph $G$.
After some calculations, we get
\begin{equation}
\sum_{\pm} \left(\frac{1}{L_{\pm}^{2}}\right)^{2} = \frac{1}{|G|}\left(\frac{1}{|G|} - \frac{2|H|}{\Delta}\right),
	\ \ \ \
\sum_{\pm} \left(\frac{\beta_{\pm}}{L_{\pm}^{2}}\right)^{2} = \frac{2}{\Delta},
\end{equation}
which completes the stated claim.
\qed\\

\ignore{
\begin{theorem}
For each $k$ and $n$, if $G = K_{1} + C$, where $C$ is a $k$-regular circulant of order $n$, 
then the continuous-time quantum walk is not uniform mixing if it starts in $K_{1}$.
\end{theorem}
\prf
By Lemma \ref{lemma:sw}, the eigenvalues of $G$ are the $n-1$ eigenvalues of $C$, say $\lambda_{1},\ldots,\lambda_{n-1}$,
except for $\lambda_{0} = k$, and $\lambda_{\pm} = \frac{1}{2}(k \pm \sqrt{k^{2} + 4n})$.
The corresponding orthogonal eigenvectors are given by
$\vec{Z}_{j} = [0, 1, \omega^{j}, \omega^{2j}, \ldots, \omega^{(n-1)j}]^{T}$, where $\omega = \exp(2\pi i/n)$,
and
$\vec{Z}_{\pm} = [x_{\pm}, 1, \ldots, 1]^{T}$, where $x_{\pm} = \frac{1}{2}(-k \pm \sqrt{k^2 + 4n})$.
Now note that $\ket{0} = [1, 0, 0, \ldots, 0]^{T}$ is a linear combination of the normalized eigenvectors of 
$\vec{Z}_{\pm}$. Let $\ell_{\pm} = \sqrt{(x_{\pm})^2 + n}$ be the length of $\vec{Z}_{\pm}$, respectively.
Thus, we have
\begin{equation}
\ket{\psi(t)} = e^{-it A_{G}} \ket{\psi(0)} 
	= \sum_{\pm} e^{-it \lambda_{\pm}} \frac{x_{\pm}}{(x_{\pm}^{2} + n)} \ \vec{Z}_{\pm} 
\end{equation}
Therefore,
\begin{equation}
\braket{0}{\psi(t)} = \sum_{\pm} e^{-it \lambda_{\pm}} \frac{x_{\pm}^{2}}{(x_{\pm}^{2} + n)}
\end{equation}
and
\begin{equation}
\overline{p}_{0} = \left(\frac{m_{+}}{m_{+} + m_{-}}\right)^2 + \left(\frac{m_{-}}{m_{+} + m_{-}}\right)^2
	= 1 - 2 \left(\frac{m_{+}}{m_{+} + m_{-}}\right)\left(\frac{m_{-}}{m_{+} + m_{-}}\right),
\end{equation}
$m_{\pm} = x_{\pm}^{2} + n$.
After further calculations, the limiting probability distribution $\overline{P}$ is given by
\begin{equation}
\overline{p}_{0} = 1 - \frac{1}{2} \left(\frac{n}{n + (k/2)^2}\right),
\ \ \
\overline{p}_{j} = \frac{1}{n}(1 - \overline{p}_{0}), \ \ j = 1,\ldots,n.
\end{equation}
This shows that on $K_{1} + K_{n}$, the quantum walk starting at the vertex of $K_{1}$ is completely
localized on $K_{1}$, while if $C$ is a sparse $k$-regular circulant, say $k = O(1)$, then the quantum walk 
has nearly equal probabilities of being on $K_{1}$ and on $C$.
\qed\\
}

\noindent The single theorem above implies the following various known and new facts about mixing on the family
of complete and related graphs. First, we obtain a perfect uniform mixing behavior on $K_{2}$, but not on $K_{n}$,
for $n > 2$.

\begin{corollary}
The continuous-time quantum walk on $K_{2}$ is average exactly uniform mixing.
\end{corollary}
\prf
By Theorem \ref{thm:join}, we have $|G| = |H| = 1$ and $k = \ell = 0$. Thus, $\Delta = 4$, and therefore,
$\overline{p}_{0} = \overline{p}_{1} = 1/2$.
\qed

\begin{corollary} \cite{aaht03}
The continuous-time quantum walk on $K_{n}$ is not average uniform mixing, as $n \rightarrow \infty$.
\end{corollary}
\prf
By Theorem \ref{thm:join}, we have $K_{n} = K_{1} + K_{n-1}$. 
We have $|G| = 1$, $|H| = n-1$, $k = 0$, and $\ell = n-1$. Then, $\Delta = (n-1)^{2} + 4n$, with
$\overline{p}_{0} = 1 - 2n/\Delta \sim 1$ and $\overline{p}_{j} = 2/\Delta \sim 0$, as $n \rightarrow \infty$.
\qed\\

\noindent Next, we consider the cone of circulants. The following corollary provides a simple explanation 
why $K_{n}$ is not average uniform mixing, for large $n$; it is because $K_{n}$ is a cone of a dense circulant.

\begin{corollary} 
The continuous-time quantum walk on the cone of any circulant $C$, namely, $K_{1} + C$, is not average uniform mixing.
\end{corollary}
\prf
Let $C$ be a $\ell$-regular circulant of order $n$. By Theorem \ref{thm:join}, we have $|G| = 1$, $|H| = n$,
$k = 0$. Then, $\overline{p}_{0} = 1 - (1/2)[1/(1 + (\ell/2)^{2}/n)]$. Thus, $\overline{p}_{0} = \Omega(1)$,
regardless of $\ell$.
\qed

\ignore{
\begin{theorem}
The random walk on the cone of any circulant is uniform mixing.
\end{theorem}
}

\paragraph{Homogeneous Joins of Circulants}
Consider the unbounded $m$-fold homogeneous join of a circulant $G$, namely, $G^{(+m)} = G + \ldots + G$, where
there are $m$ terms in the summation. The following theorem shows that the uniform mixing property of $G$ transfers 
into its unbounded homogeneous join if $m$ is a constant.

\begin{theorem}
Let $G$ be a circulant of order $n$. Let $m \ge 2$ is a constant and $n > 2\lambda_{0}(G)$.
In the continuous-time quantum walk, $G^{(+m)} = \sum_{\ell=1}^{m} G$ is average uniform mixing if $G$ is.
\end{theorem}
\prf
The adjacency matrix of $G^{(+m)} = \sum_{\ell=1}^{m} G$ is given by
\begin{equation}
A = I_{m} \otimes G + K_{m} \otimes J_{n},
\end{equation}
where $I_{m}$ is the $m \times m$ identity matrix, $K_{m}$ is a complete graph on $m$ vertices,
and $J_{n}$ is the $n \times n$ all-one matrix. Since $G$ is a circulant, both summands share the same
set of the following orthonormal eigenvectors
\begin{equation}
\left\{\ket{F_{j,k}} = \ket{F^{(m)}_{j}} \otimes \ket{F^{(n)}_{k}} \ : \ 0 \le j \le m-1, 0 \le k \le n-1\right\},
\end{equation}
where $\ket{F^{(m)}_{j}}$ denotes the $j$-th column of the $m \times m$ Fourier matrix, and
similary for $\ket{F^{(n)}_{k}}$. Let $\lambda_{k}(G)$, for $0 \le k \le n-1$, be the eigenvalues of $G$ in
descending order. The corresponding eigenvalues of $G^{(+m)}$ are given by
\begin{equation}
\lambda_{j,k} = 
	\left\{\begin{array}{ll}
	\lambda_{0}(G) + (m-1)n		& \mbox{ if $j = k = 0$ } \\
	\lambda_{0}(G) - n		& \mbox{ if $j \neq 0$ and $k = 0$ } \\
	\lambda_{k}(G)			& \mbox{ if $j,k \neq 0$ }
	\end{array}\right.
\end{equation}
If $\ket{\psi(0)} = \ket{0} \otimes \ket{0}$ then
$\ket{\psi(0)} = 1/\sqrt{mn}\sum_{j=0}^{m-1}\sum_{k=0}^{n-1} \ket{F_{j,k}}$. Thus,
\begin{equation}
\ket{\psi(t)} = \frac{1}{\sqrt{mn}}\sum_{j=0}^{m}\sum_{k=0}^{n-1} e^{-it\lambda_{j,k}} \ket{F_{j,k}}.
\end{equation}
Thus, for $x \in \Int_{m}$ and $y \in \Int_{n}$, we have
\begin{equation}
\psi_{x,y}(t) 
	= \braket{x,y}{\psi(t)} 
	= \frac{1}{mn}\sum_{j=0}^{m}\sum_{k=0}^{n-1} e^{-it\lambda_{j,k}} 
		\exp\left(\frac{2\pi ijx}{m}\right) 
		\exp\left(\frac{2\pi iky}{n}\right).
\end{equation}
Note that the three types of eigenvalues of $G^{(+m)}$ are mutually distinct, since
\begin{equation}
\lambda_{0}(G) - n \ < \ \lambda_{k}(G) \ < \ \lambda_{0}(G) + (m-1)n.
\end{equation}
\ignore{
\begin{equation}
\overline{p}_{x,y} 
	= \frac{1}{mn} 
	+ \frac{1}{(mn)^{2}}\sum_{(j_1,k_1) \neq (j_2,k_2)} e^{-it(\lambda_{j_1,k_1}-\lambda_{j_2,k_2})}
		\exp\left(\frac{2\pi i(j_1 - j_2)x}{m}\right) \exp\left(\frac{2\pi i(k_1 - k_2)y}{n}\right).
\end{equation}
}
Therefore, we have
\begin{equation}
\left|\overline{p}_{x,y} - \frac{1}{mn}\right|
	\leq \frac{1}{(mn)^{2}} \ \binom{m-1}{2} \left[1 + n\binom{\mu(G)}{2}\right]
	= O\left(\frac{\mu^{2}(G)}{mn}\right),
\end{equation}
since $m$ is a constant.
\qed\\

\par\noindent
The above theorem also {\em explains} why the complete graph $K_{N}$ is not uniform mixing, since 
$K_{N}$ can be viewed as a homogeneous $m$-fold join of $K_{N/m}$, for some constant $m$ that divides $N$.
The theorem also implies the following claim about the multipartite complete graphs.

\begin{corollary}
The continuous-time quantum walk on the complete multipartite graph $K^{(m)}_{n}$
is not average uniform mixing if $m \ge 2$ is a constant.
\end{corollary}
\prf
Since a continuous-time quantum walk is not average uniform mixing on the empty graph $\overline{K}_{n}$ 
and $K^{(m)}_{n} = \overline{K}_{n} + \ldots + \overline{K}_{n}$, we have our claim.
\qed


\section{Mixing on Cartesian Bunkbeds}

In this section, we consider a circulant bunkbed structure obtained by the Cartesian product 
$P_{2} \oplus C$, where $C$ is a circulant graph. 

\begin{lemma} \label{lemma:plus}
Let $G$ be a circulant of degree $d$ and order $n$, whose eigenvalues are 
$d = \mu_{0} > \mu_{1} \ge \ldots \ge \lambda_{n-1}$. Then, the eigenvalues of $P_{2} \oplus G$
are $\lambda^{\pm}_{j} = \mu_{j} \pm 1$ with the following (orthonormal) set of eigenvectors
\begin{equation}
\ket{z^{\pm}_{j}} = \ket{\pm} \otimes \ket{z_{j}},
\end{equation}
where $\ket{\pm} = \frac{1}{\sqrt{2}}(\ket{0} \pm \ket{1})$ 
and $\braket{x}{z_{j}} = (1/\sqrt{|G|}) \omega^{jx}$, for $j,x \in [n]$, with $\omega = e^{2\pi i/n}$.
\end{lemma}
\prf
Note that the adjacency matrix of $P_{2} \oplus G$ is given by $P_{2} \otimes I_{n} + I_{2} \otimes A_{G}$.
Since $P_{2}$ is a circulant, both $P_{2} \otimes I_{n}$ and $I_{2} \otimes A_{G}$ are simultaneously
diagonalizable by $\ket{z^{\pm}_{j}}$. This implies the stated claim on the spectra of $P_{2} \oplus G$.
\qed

\ignore{
\begin{figure}[h]
\[\begin{xy} /r10mm/:
 ,{\xypolygon4"A"{~={0}\dir{*}}}
 ,+(0.0,0.75),{\xypolygon4"B"{~={0}\dir{*}}}
 ,"A1";"B1"**@{-},"A2";"B2"**@{-},"A3";"B3"**@{-},"A4";"B4"**@{-}
 ,"A1";"B2"**@{-},"A2";"B3"**@{-},"A3";"B4"**@{-},"A4";"B1"**@{-}
 ,"A1";"B3"**@{-},"A2";"B4"**@{-},"A3";"B1"**@{-},"A4";"B2"**@{-}
 ,"A1";"B4"**@{-},"A2";"B1"**@{-},"A3";"B2"**@{-},"A4";"B3"**@{-}
 ,+/r40mm/
 ,{\xypolygon4"M"{~={0.25}\dir{*}}}
 ,+(0.0,0.75),{\xypolygon4"N"{~={0.25}\dir{*}}}
 ,"M1";"N1"**@{-},"M2";"N2"**@{-},"M3";"N3"**@{-},"M4";"N4"**@{-}
\end{xy}\]
\caption{Some circulant bunkbed.
}
\label{figure:circulant-bunkbed}
\end{figure}
}

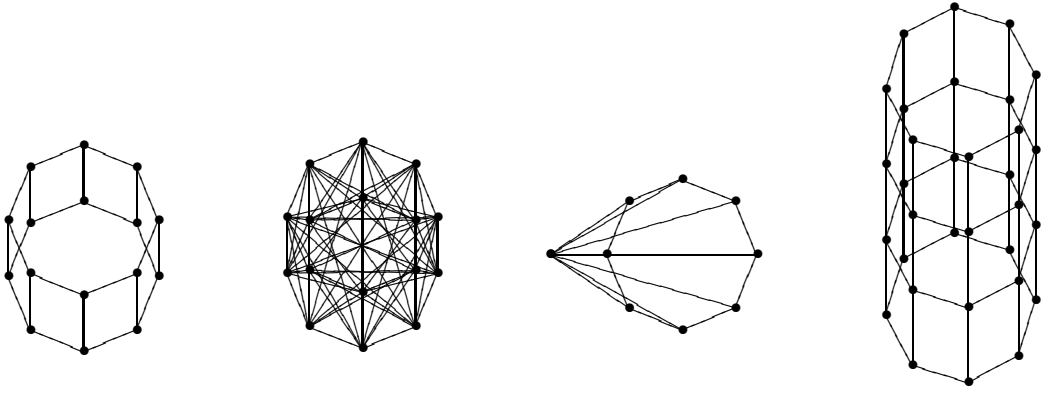
\begin{figure}[t]
\[\begin{xy} /r10mm/:
 ,{\xypolygon8"P"{~={0}\dir{*}}}
 ,+(0.0,0.75),{\xypolygon8"Q"{~={0}\dir{*}}}
 ,"P1";"Q1"**@{-},"P2";"Q2"**@{-},"P3";"Q3"**@{-},"P4";"Q4"**@{-}
 ,"P5";"Q5"**@{-},"P6";"Q6"**@{-},"P7";"Q7"**@{-},"P8";"Q8"**@{-}
 ,+/r30mm/
 ,{\xypolygon8"A"{~={0}\dir{*}}}
 ,+(0.0,0.75),{\xypolygon8"B"{~={0}\dir{*}}}
 ,"A1";"B1"**@{-},"A2";"B2"**@{-},"A3";"B3"**@{-},"A4";"B4"**@{-}
 ,"A5";"B5"**@{-},"A6";"B6"**@{-},"A7";"B7"**@{-},"A8";"B8"**@{-}
 ,"A1";"B2"**@{-},"A2";"B3"**@{-},"A3";"B4"**@{-},"A4";"B5"**@{-}
 ,"A5";"B6"**@{-},"A6";"B7"**@{-},"A7";"B8"**@{-},"A8";"B1"**@{-}
 ,"A1";"B3"**@{-},"A2";"B4"**@{-},"A3";"B5"**@{-},"A4";"B6"**@{-}
 ,"A5";"B7"**@{-},"A6";"B8"**@{-},"A7";"B1"**@{-},"A8";"B2"**@{-}
 ,"A1";"B4"**@{-},"A2";"B5"**@{-},"A3";"B6"**@{-},"A4";"B7"**@{-}
 ,"A5";"B8"**@{-},"A6";"B1"**@{-},"A7";"B2"**@{-},"A8";"B3"**@{-}
 ,"A1";"B5"**@{-},"A2";"B6"**@{-},"A3";"B7"**@{-},"A4";"B8"**@{-}
 ,"A5";"B1"**@{-},"A6";"B2"**@{-},"A7";"B3"**@{-},"A8";"B4"**@{-}
 ,"A1";"B6"**@{-},"A2";"B7"**@{-},"A3";"B8"**@{-},"A4";"B1"**@{-}
 ,"A5";"B2"**@{-},"A6";"B3"**@{-},"A7";"B4"**@{-},"A8";"B5"**@{-}
 ,"A1";"B7"**@{-},"A2";"B8"**@{-},"A3";"B1"**@{-},"A4";"B2"**@{-}
 ,"A5";"B3"**@{-},"A6";"B4"**@{-},"A7";"B5"**@{-},"A8";"B6"**@{-}
 ,"A1";"B8"**@{-},"A2";"B1"**@{-},"A3";"B2"**@{-},"A4";"B3"**@{-}
 ,"A5";"B4"**@{-},"A6";"B5"**@{-},"A7";"B6"**@{-},"A8";"B7"**@{-}
 ,+/r15mm/
 ,+(0.0,0.5),
 ,{\xypolygon1"P"{~={0}\dir{*}}}
 ,+(2.75,0.0),{\xypolygon8"Q"{~={0}\dir{*}}}
 ,"P1";"Q1"**@{-},"P1";"Q2"**@{-},"P1";"Q3"**@{-},"P1";"Q4"**@{-}
 ,"P1";"Q5"**@{-},"P1";"Q6"**@{-},"P1";"Q7"**@{-},"P1";"Q8"**@{-}
 ,+/r30mm/
 ,{\xypolygon8"P"{~={50}\dir{*}}}
 ,+(0.0,1.0),{\xypolygon8"Q"{~={50}\dir{*}}}
 ,+(0.0,1.0),{\xypolygon8"R"{~={50}\dir{*}}}
 ,+(0.0,1.0),{\xypolygon8"S"{~={50}\dir{*}}}
 ,"P1";"Q1"**@{-},"P2";"Q2"**@{-},"P3";"Q3"**@{-},"P4";"Q4"**@{-}
 ,"P5";"Q5"**@{-},"P6";"Q6"**@{-},"P7";"Q7"**@{-},"P8";"Q8"**@{-}
 ,"R1";"Q1"**@{-},"R2";"Q2"**@{-},"R3";"Q3"**@{-},"R4";"Q4"**@{-}
 ,"R5";"Q5"**@{-},"R6";"Q6"**@{-},"R7";"Q7"**@{-},"R8";"Q8"**@{-}
 ,"R1";"S1"**@{-},"R2";"S2"**@{-},"R3";"S3"**@{-},"R4";"S4"**@{-}
 ,"R5";"S5"**@{-},"R6";"S6"**@{-},"R7";"S7"**@{-},"R8";"S8"**@{-}
\end{xy}\]
\caption{Examples of circulant bunkbeds.
(a) The double bunkbed: $P_{2} \oplus C_{8}$.
(b) The Join bunkbed: $C_{8} + C_{8}$.
(c) The Cone: $K_{1} + C_{8}$.
(d) The Cartesian bunkbed: $P_{4} \oplus C_{8}$.
}
\label{figure:circulant-bunkbed}
\end{figure}

\begin{theorem}
Let $G$ be a circulant of order $n$. In the continuous-time quantum walk,
$P_{2} \oplus G$ is average uniform mixing if $G$ is.
\end{theorem}
\prf
Assume that $\ket{\psi(0)} = \ket{0} \otimes \ket{0}$. 
Thus, $\ket{\psi(0)} = \sum_{\pm} \frac{1}{\sqrt{2}} \ket{\pm} \otimes \sum_{j} \frac{1}{\sqrt{n}} \ket{z_{j}}$, and
\begin{equation}
\ket{\psi(t)} 
	= e^{-it A} \ket{\psi(0)}
	= \frac{1}{\sqrt{2n}} \sum_{\pm,j} e^{-it \lambda^{\pm}_{j}} \ket{\pm} \otimes \ket{z_{j}}.
\end{equation}
This implies that
\begin{eqnarray}
\braket{b,x}{\psi(t)}
	& = & \frac{1}{\sqrt{2n}} \sum_{\pm,j} e^{-it(\lambda_{j}\pm 1)} \braket{b}{\pm} \braket{x}{z_{j}} \\
	& = & \frac{1}{2n} \sum_{\pm,j} e^{-it(\lambda_{j}\pm 1)} (\pm 1)^{b} \omega^{jx} \\
	& = & \frac{1}{n} \sum_{j} e^{-it\lambda_{j}} \omega^{jx} \sum_{\pm} e^{-it(\pm 1)} (\pm 1)^{b} \\
	& = & \frac{1}{n} \sum_{j} e^{-it\lambda_{j}} \omega^{jx} [(1-b)\cos(t) + b(-i\sin(t))]
\end{eqnarray}
Let $p_{b,x}(t) = |\braket{b,x}{\psi(t)}|^{2}$. Thus,
\begin{equation}
p_{b,x}(t) = \frac{1}{n^{2}} \sum_{j,k} [(1-b)\cos^{2}(t) + b\sin^{2}(t)]
		e^{-it(\lambda_{j} - \lambda_{k})} \omega^{(j-k)x}.
\end{equation}
Note that $p_{0,x}(t) + p_{1,x}(t) = p_{x}(t)$, where $p_{x}(t)$ is the (instantaneous) probability 
on vertex $x$ at time $t$ of a quantum walk on $G$ alone. Then,
\begin{equation}
\overline{p}_{0,x} 
	= \frac{1}{n^{2}} \sum_{j,k} \omega^{(j-k)x} 
		\lim_{T \rightarrow \infty} \frac{1}{T} 
		\int_{0}^{T} \cos^{2}(t) e^{-it(\lambda_{j}-\lambda_{k})} \ \dt
	= \frac{1}{2} \ \overline{p}_{x},
\end{equation}
since $\lim_{T \rightarrow \infty} \frac{1}{T} \int_{0}^{T} \cos^{2}(t) e^{-it\Delta} \ \dt =
\frac{1}{2} \ \iverson{\Delta = 0}$. Similarly, we obtain $\overline{p}_{1,x} = \frac{1}{2}\overline{p}_{x}$.
This yields the claim.
\qed

\begin{corollary}
The continuous-time quantum walk on a Cartesian bunkbed $P_{2} \oplus G$, 
where $G$ is a $d$-degree circulant of the form $\langle 1, n/k_{1}, \ldots, n/k_{d-1}\rangle$, 
where $d$ and $k_{1},\ldots,k_{d-1}$ are constants, is average uniform mixing.
\end{corollary}

\paragraph{Circulant Cylinders}
To extend our Cartesian bunkbeds over paths with more than two vertices, we provide, for completeness, 
an analysis of the quantum walk on paths. This problem is well-known in the physics literature, 
but is normally done on the infinite paths using different techniques \cite{fls}.
The eigenvalues $\lambda_{j}$ and eigenvectors $\ket{Q_{j}}$ of the path $P_{m}$ (see \cite{spitzer}),
for $j=1,\ldots,m$, are defined as
\begin{eqnarray}
\lambda_{j} & = & 2\cos\left(\frac{j\pi}{m+1}\right) \\
\braket{x}{Q_{j}} & = & \frac{1}{\sqrt{(m+1)/2}} \ \sin\left(\frac{jx \pi}{m+1}\right),	\ \ \  x = 1,\ldots,m
\end{eqnarray}
If the quantum walk starts with the initial state $\ket{\psi(0)} = \ket{1}$, where the basis states
are $\ket{1}, \ldots, \ket{m}$, then
\ignore{
\begin{equation}
\ket{\psi(0)} = \sum_{j=1}^{m} \frac{1}{\sqrt{(m+1)/2}} \ \sin\left(\frac{j\pi}{m+1}\right) \ \ket{Q_{j}}.
\end{equation}
}
Thus, we have
\begin{equation}
\ket{\psi(t)} = 
\sum_{j=1}^{m} \frac{e^{-it\lambda_{j}}}{\sqrt{(m+1)/2}} \ \sin\left(\frac{j\pi}{m+1}\right) \ \ket{Q_{j}}.
\end{equation}
\ignore{
and
\begin{equation}
\braket{x}{\psi(t)} = 
\frac{2}{m+1}
\sum_{j=1}^{m} e^{-it\lambda_{j}} \ \sin\left(\frac{j\pi}{m+1}\right) \ \sin\left(\frac{jx\pi}{m+1}\right)
\end{equation}
}
Since $P_{m}$ has $m$ distinct eigenvalues, the limiting probabilities are given by
\begin{equation}
\overline{p}_{x} = \frac{4}{(m+1)^2} \sum_{j=1}^{m} \ \sin^{2}\left(\frac{j\pi}{m+1}\right)
	\sin^{2}\left(\frac{jx\pi}{m+1}\right).
\end{equation}
Note that, since $\int_{0}^{\pi} \sin^2(t) \dt = \pi/2$, we get an upper bound of
\begin{eqnarray} 
\overline{p}_{x} 
	& \leq & \frac{4}{(m+1)^2} \sum_{j=1}^{m} \ \sin^{2}\left(\frac{j\pi}{m+1}\right) \\
	& \leq & \frac{4}{(m+1)\pi} \left(\int_{0}^{\pi} \sin^{2}(t) \ \dt \ + \ \frac{\pi}{(m+1)}\right) \\
	& \leq & \frac{2}{(m+1)} + \frac{4}{(m+1)^2} = O\left(\frac{1}{m}\right),
\end{eqnarray}
which implies that the quantum walk on $P_{m}$ is average uniform mixing.
\ignore{
and, since $\sin(x) \ge x - x^{3}/6 \ge x/2$, we get a lower bound of 
\begin{equation} \label{eqn:lower}
\overline{p}_{x} 
	\ge \frac{4}{(m+1)^2} \sum_{j=1}^{m} \ \sin^{2}\left(\frac{j\pi}{m+1}\right) \frac{\pi^{2}}{4(m+1)^2}
\end{equation}
}

The eigenvalues of a circulant cylinder $T = P_{m} \oplus G$, where $G$ is a circulant of order $n$, are given by
\begin{equation}
\lambda_{j,k} = \mu_{j} + \nu_{k}, \ \ \ \ \ \mbox{ where } \ \ 1 \le j \le m, \ \ 0 \le k \le n-1,
\end{equation}
where $\mu_{j} = 2\cos(j\pi/(m+1))$ and $\nu_{k}$ are the eigenvalues of $P_{m}$ and $G$, respectively.
Since the adjacency matrix of $T$ is defined as $P_{m} \otimes I_{n} + I_{m} \otimes G$,
the eigenvectors of $T$ are 
\begin{equation}
\ket{T_{j,k}} = \ket{Q_{j}} \otimes \ket{F_{k}}, \ \ \ \ \ \mbox{ where } \ \ 1 \le j \le m, \ \ 0 \le k \le n-1,
\end{equation}
where $\ket{Q_{j}}$ and $\ket{F_{j}}$ are the eigenvectors of $P_{m}$ and the circulant $G$, respectively.
Recall that $\braket{x}{Q_{j}} = \sqrt{\frac{2}{m+1}} \sin\left(\frac{jx\pi}{m+1}\right)$, for $1 \le j,x \le m$, 
and $\braket{y}{F_{k}} = \frac{1}{\sqrt{n}} \exp\left(\frac{2\pi i ky}{n}\right)$, for $0 \le k,y \le n-1$.

If the initial state is $\ket{\psi(0)} = \ket{1} \otimes \ket{0}$, we have
\begin{equation}
\ket{\psi(0)} = 
	\sum_{j=1}^{m} \braket{Q_{j}}{1} \ket{Q_{j}} 
	\otimes 
	\sum_{k=0}^{n-1} \braket{F_{k}}{0} \ket{F_{k}}.
\end{equation}
The adjacency matrix of $P_{m} \oplus G$ is given by $A = P_{m} \otimes I_{n} + I_{m} \otimes G$, where the two
summands commute with each other. Thus, $e^{-it A} = e^{-it(P_{m} \otimes I_{n})} e^{-it(I_{m} \otimes G)}$, and
\begin{equation}
\ket{\psi(t)} = 
	\sum_{j=1}^{m} \braket{Q_{j}}{1} e^{-it\mu_{j}} \ket{Q_{j}} 
	\otimes 
	\sum_{k=0}^{n-1} \braket{F_{k}}{0} e^{-it\nu_{k}} \ket{F_{k}}.
\end{equation}
The amplitudes of $\ket{\psi(t)}$ at vertex $x$ on the path $P_{m}$ and vertex $y$ within the circulant $G$ is given by
\begin{eqnarray}
\braket{x,y}{\psi(t)} 
	& = &
	\sum_{j=1}^{m} e^{-it\mu_{j}} \braket{x}{Q_{j}} \braket{Q_{j}}{1} 
	\sum_{k=0}^{n-1} e^{-it\nu_{k}} \braket{y}{F_{k}} \braket{F_{k}}{0} 
\end{eqnarray}

\begin{corollary}
Let $G$ be a Cartesian product $P_{m} \oplus C$, where $C$ is a circulant of order $n$.
The continuous-time quantum walk on $G$ is uniform mixing if $m$ is constant or $n$ is constant.
\end{corollary}

\ignore{
\begin{figure}[h]
\[\begin{xy} /r10mm/:
 ,{\xypolygon8"P"{~={50}\dir{*}}}
 ,+(0.0,1.0),{\xypolygon8"Q"{~={50}\dir{*}}}
 ,+(0.0,1.0),{\xypolygon8"R"{~={50}\dir{*}}}
 ,+(0.0,1.0),{\xypolygon8"S"{~={50}\dir{*}}}
 ,"P1";"Q1"**@{-},"P2";"Q2"**@{-},"P3";"Q3"**@{-},"P4";"Q4"**@{-}
 ,"P5";"Q5"**@{-},"P6";"Q6"**@{-},"P7";"Q7"**@{-},"P8";"Q8"**@{-}
 ,"R1";"Q1"**@{-},"R2";"Q2"**@{-},"R3";"Q3"**@{-},"R4";"Q4"**@{-}
 ,"R5";"Q5"**@{-},"R6";"Q6"**@{-},"R7";"Q7"**@{-},"R8";"Q8"**@{-}
 ,"R1";"S1"**@{-},"R2";"S2"**@{-},"R3";"S3"**@{-},"R4";"S4"**@{-}
 ,"R5";"S5"**@{-},"R6";"S6"**@{-},"R7";"S7"**@{-},"R8";"S8"**@{-}
\end{xy}\]
\caption{The cylindrical circulant bunkbed $P_{4} \oplus C_{8}$.}
\label{figure:cartesian-bunkbed}
\end{figure}
}


\section{Conclusions}

It was known that a continuous-time quantum walk is uniform average mixing on the cycles $C_{n}$,
but is not uniform average mixing on the complete graphs $K_{n}$ and on the hypercubes $Q_{n}$. 
Our goal in this work was to provide a graph-theoretic explanation for this polarized phenomena.

First, we extend the phenomenon of the cycles, by showing that uniform mixing is achieved on circulants 
with bounded eigenvalue multiplicity. We also gave other explicit examples of circulants meeting this 
criteria. Second, we consider two graph-theoretic bunkbed structures over circulants in order to study 
the non-uniform mixing on $K_{n}$ and $Q_{n}$. Our analysis on the join bunkbed sheds some light on the 
non-uniform mixing of the complete multipartite graphs (which includes $K_{n}$). Our analysis of the 
Cartesian bunkbed of circulants highlights a difference between the $\Int_{n}$-circulants and the 
$(\Int_{2})^{n}$-circulants (see \cite{d90}). We leave a similar investigation of general group-theoretic 
circulants and Cayley graphs for future work (see \cite{gw03}).

\ignore{
We conclude with the following open questions:
\begin{itemize}
\item Do all bounded-degree circulants have bounded eigenvalue multiplicities?
\item Are the cycles unique in being the only circulants having $\mu(G) = 2$?
\item What is the explanation for the non-uniform mixing phenomenon on the hypercubes?
\end{itemize}
}

\ignore{
\appendix

\section{Circulant Cylinders}

\begin{theorem}
The hitting time of vertex $t$ of a random walk on $G$ that starts at $s$ is
\begin{equation}
\frac{1}{2}(m-1)(2 + (k + 2)(m - 2)) = O(k m^{2}).
\end{equation}
\end{theorem}

\begin{conjecture}
The hitting time of vertex $t$ of a quantum walk on $G$ that starts at $s$ is $O(km)$.
\end{conjecture}
}

\end{document}